\documentclass[showpacs,preprintnumbers,amsmath,amssymb,aps]{revtex4}

\begin{document}

\title{Comment on \textquotedblleft Temperature dependent fluctuations in
the two-dimensional XY model\textquotedblright }
\author{G. Palma \\
Departamento de F\'{\i}sica, Universidad de Santiago de Chile, \\
Casilla 307, Santiago 2, Chile} \pacs{05.70.Jk, 05.40.-a,
05.50,75.10.Hk, 05.50.+q} \maketitle

In \cite{PML}, the temperature dependence of the probability density
function (PDF) of the magnetization (the order parameter) was first shown by
numerical Monte Carlo analysis using the finite 2D XY model. This result
contradicted the previous claims found in the literature \cite{BFHPPPS}.
Recently, \cite{BB} shows semi-analytically that the PDF depends on the
system temperature $T$. To show this dependence the authors compute the
contribution of the multi loop graphs (MLG)\ to the moments of the PDF
within the Harmonic approximation of the XY model. They show that the MLG
depend on $T$ and may not be neglected in the thermodynamic limit contrary
to previous assumptions \cite{BFHPPPS}. Using a Monte Carlo simulation they
compute the skewness $\gamma _{3}$ or normalized third moment of the PDF,
which \textquotedblleft provides a clear measure of the variation of the PFD
with temperature\textquotedblright , and using a Monte Carlo simulation,\
obtain the numerically approximated expression for $\gamma _{3}$, valid for
a square lattice of lattice size $L=16$,

\begin{equation*}
\gamma _{3}(T)\ \approx -0.85+0.126T-0.0048T^{2}\
\end{equation*}%
They also obtained for the lattice size $L=32$ \textquotedblleft a value
much closer to the theory\textquotedblright\ with $\gamma _{3}(T)\ \approx
-0.88+0.15T$, and finally argued that the skewness is relatively
computational expensive due to the need for averaging and, as their results
appear to confirm the evolution of $\gamma _{3}$ with $T$, they
\textquotedblleft ... leave the determination of the precise form of $\gamma
_{3}(T)$ from larger systems to another time\textquotedblright .

This comment is devoted first to show that an explicit analytical expression
for $\gamma _{3}(T)$, valid for arbitrary system size and to all orders in $%
T $, can be deduced starting from their equation (11) for the higher moments
$\langle m^{p}\rangle $ of the PDF, and second to show that the numerical
values of the slope of $\gamma _{3}$ reported in their paper are not just
numerically incorrect, but their scaling with the lattice size is wrong.

Starting from their equation (11) and using the translational invariance of
the lattice propagator $G$, the exact expression for the moments to all
orders in $T$ can be obtained, yielding:\bigskip
\begin{equation}
\left\langle \ M\ \right\rangle =\exp [-T\ G(0)\ /\ 2]  \label{m1}
\end{equation}%
\begin{equation}
\left\langle ~M^{2}\ \right\rangle \ =\ \frac{\left\langle \ M\
\right\rangle ^{2}}{N}\sum\limits_{\overrightarrow{x}\epsilon
\Lambda }\cosh [T(G(\overrightarrow{x})]  \label{m2}
\end{equation}%
\begin{eqnarray}
\left\langle ~M^{3} \right\rangle  = \frac{\left\langle  M
\right\rangle ^{3}}{2\ N^{2}}\sum\limits_{\overrightarrow{x},%
\overrightarrow{y}\epsilon \Lambda }\bigg\{ \exp (-T G(\overrightarrow{x}%
)) \cosh [T(G(\overrightarrow{y})+G(\overrightarrow{x}-%
\overrightarrow{y})]+\exp ( T G(%
\overrightarrow{x})) \cosh [T ( G(\overrightarrow{y})-G (\overrightarrow{%
x}-\overrightarrow{y})]\bigg\}
\end{eqnarray}%
where $\Lambda $ denotes the lattice, $N=L^{2}$ is the volume and the
lattice propagator $G$ is given for example by its Fourier representation
\begin{equation*}
G(\overrightarrow{x})=\frac{1}{N}\sum\limits_{(\overrightarrow{K}_{L}\
)^{2}\neq 0}\frac{\exp (-i\overrightarrow{K}\cdot \overrightarrow{x})}{(%
\overrightarrow{K_{L}})^{2}}\
\end{equation*}%
$\overrightarrow{K_{L}}$ is the lattice momentum defined as usual as $%
(K_{L})_{i}\ =2\sin (K_{i}/2)$ with $i=1,2$ and $K_{i}$ lies in the first
Brillouin zone. We now use the definition of the skewness as the third
normalized moment $\gamma _{3}(T)=\left\langle \ [(M-\langle \ M\ \rangle \
)\ /\ \sigma \ ]^{3}\right\rangle $ and write%
\begin{equation}
\gamma _{3}(T)=\frac{1}{\left\{ \left\langle ~M^{2}\ \right\rangle
-\left\langle ~M\ \right\rangle ^{2}\right\} ^{3/2}}\left[ \left\langle
~M^{3}-3\left\langle ~M^{2}\right\rangle \left\langle ~M\ \right\rangle
+2\left\langle ~M\ \right\rangle ^{3}\ \right\rangle \right]   \label{mu3}
\end{equation}%
and explicitly expanding up to order $T$, we obtain
\begin{equation}
\gamma _{3}(T)=-g_{3}\left( \frac{2}{g_{2}}\right) ^{3/2}\left\{ 1-\frac{3}{4%
}\frac{(g_{2})^{2}}{g_{3}}\ T\ +\ O(T^{2})\right\}   \label{skewness}
\end{equation}%
where the quantities $g_{n}$ are defined in terms of the power $n$ of the
lattice propagator $G$ as $g_{n}=G^{n}(0)/N^{n-1}$.\ The expression of eqn. (%
\ref{skewness}) agrees with the corresponding equation (33) of ref. \cite%
{MPV}, where the skewness and kurtosis are computed for the full 2D XY-model
up to two-loops, including the anharmonic corrections to the Hamiltonian
(see the two last terms in eqn. (33)), which are suppressed by a volume
factor $N$, and which are negligible in the thermodynamic limit but are
relevant for finite lattice sizes. Higher order corrections in $T$ can be
directly computed along the lines outlined here. From the numerical point of
view, we use the known numerical values for the lattice coefficients $g_{n}$%
, which are directly and easily evaluated by using MATLAB for example, and
obtain $\gamma _{3}(T)\ \approx -0.8540+0.1358T$ for $L=16$ and $\gamma
_{3}(T)\ \approx -0.8763+0.1331T$ for $L=32$ respectively. Independent of
the numerical differences found for the first two coefficients of the
skewness, one observes that the values for the slope (linear term in $T$)
reported in \cite{BB} increases with the system size, contrary to the
tendency of the analytic expression displayed by eqn. (\ref{skewness}). In
fact, the slope is a decreasing function of the system size and its
thermodynamic limit converges to the value 0.1319. Finally, and in order to
compare our results with the accurate numerical values obtained in \cite{MPV}
for the skewness, we evaluate the contribution obtained from the anharmonic
corrections

\begin{equation*}
\delta \gamma _{3}(T)=\frac{3}{2N}\ g_{3}~\left( \frac{2}{g_{2}}\right)
^{3/2}\left( \frac{g_{1}^{2}}{2g_{2}}-\frac{g_{1}g_{2}}{g_{3}}\right) T
\end{equation*}%
and obtain $\delta \gamma _{3}(T)\approx 0.0197T$ for $L=16$ and $\delta
\gamma _{3}(T)\approx 0.0191T$ for $L=32$ respectively . Perfect agreement
is found with the corresponding value in \cite{MPV}, (see eqn.( 36)), when
adding this contribution to the value obtained in eqn. (\ref{skewness}) for $%
\gamma _{3}(T).$

\bigskip

\bigskip \bigskip This work was partially supported by FONDECYT N%
${{}^o}$
1050266. I like to thank R. Labb\'{e} and L. Vergara for valuable
discussions.

\end{document}